\documentclass{PoS}

\def \be {\begin{equation}}
\def \ee {\end{equation}}
\def \bea {\begin{eqnarray}}
\def \eea {\end{eqnarray}}
\def \nn {\nonumber}

\def \e {{\rm e}}

\def \ep{\epsilon}
\def \vep{\varepsilon}

\def \lab #1 {\label{#1}}

\def \as {{\alpha_s}}

\def \bbeta {\bar\beta}

\definecolor{agr}{rgb}{0.0,0.4,0.0}
\definecolor{apu}{rgb}{0.3,0.2,0.6}
\definecolor{dre}{rgb}{0.6,0.0,0.1}

\ifx\pdfoutput\undefined
\input epsf

\def\figscale#1#2{\epsfxsize=#2\epsfbox{#1.eps}}
\else

\def\figscale#1#2{\pdfximage width#2 {#1.pdf}\pdfrefximage\pdflastximage}
\fi

\title{A coordinate description of partonic processes}

\ShortTitle{A coordinate description of partonic processes}

\author{\speaker{George Sterman}\\
       C.N.\ Yang Institute for Theoretical Physics and Department of Physics and Astronomy, \\ Stony Brook University, Stony Brook, NY 11794-3840, USA\\
       E-mail: \email{george.sterman@stonybrook.edu}}

\author{Ozan Erdo\u{g}an \\
        Cavendish~Laboratory, University~of~Cambridge, Cambridge~CB3~0HE, UK and \\
        Department of Physics, Carnegie Mellon University, Pittsburgh, PA 15213, USA \\
        E-mail: \email{erdogan@cmu.edu}}

\abstract{We review a perturbative description of amplitudes in coordinate space, aiming at an intuitive perspective on the roles of ``long"  and ``short" distances.    We begin with coordinate-space leading regions for fixed-angle scattering amplitudes, and develop approximations and factorizations in analogy to similar procedures in momentum space.   There are also applications to products of Wilson lines, including the familiar cusp amplitude.    We briefly discuss cross sections from the coordinate viewpoint, and the mechanism by which infrared singularities cancel in inclusive cross sections.}

\FullConference{12th International Symposium on Radiative Corrections (Radcor 2015) and LoopFest XIV (Radiative Corrections for the LHC and Future Colliders)\\
		 15-19 June  2015\\
		 UCLA Department of Physics \& Astronomy Los Angeles, CA, USA}

\begin{document}

\section{Introduction: Coordinate-space leading regions}  

The all-orders analysis of long-distance behavior of perturbative scattering amplitudes in terms of loop momenta (see, for example, \cite{Sterman:1978bi}-\cite{Feige:2014wja})
 complements fixed-order calculations, and helps explain their underlying structure, especially the origins of infrared (long distance) sensitivity, associated with collinear and soft momentum configurations.  
In this talk, we'll describe how long distance sensitivity arises when viewed directly in coordinate space \cite{Date:1982un}-\cite{Erdogan:2014gha}.

We begin with Green functions,
\bea
G_N(x_1, \dots, x_N)\ &=&\ \langle 0| T\left( \phi_N(x_N)\, \cdots \ \phi_1(x_1) \right) |0\rangle\, ,
\label{eq:VEV}
\eea
 with massless fields $\phi_I$ at positions $x_I$ that reflect the geometry of scattering amplitudes.  In perturbation theory, these Green functions are sums of diagrams, which themselves can be written as sums of integrals over the positions of internal vertices $y_k$, very schematically,
\bea
G_N(x_1, \dots, x_N)\ &=&\ 
\sum_{\rm diagrams}\ \prod_{\mathrm{vertices}\ k}\int d^Dy_k
\prod_{\mathrm{lines}\ j}\,
\frac{\mathrm{numerator}}{\left [-(\sum_{k'}\eta_{jk'}\,y_{k'}+\sum_{l}\eta_{jl}\,x_l)^2+i\epsilon \right]^{p_j}}\, ,
\label{eq:GN}
\eea
where the $\eta_{ji}$ are incidence matrices.
The nontrivial numerators include overall factors and in theories with spin, derivatives acting on the propagators in coordinate space.   The basic, scalar propagator is given in coordinate space, with $D=4-2\vep$, by
\bea
\Delta(y-x)\ =\  \frac{\Gamma(1-\vep)}{4\pi^{2-\vep}}\ \frac{1}{\left( -(y-x)^2+i\ep\right)^{1-\vep}} \ . 
\label{eq:scalarprop}
\eea
More generally, in Eq.\ (\ref{eq:GN}),  $p_j = 1-\vep$ for a boson propagator (choosing Feynman gauge where necessary) and $ 2-\vep$ for a fermion, or derivative of a boson propagator.   

\begin{figure}
 \centerline{ \figscale{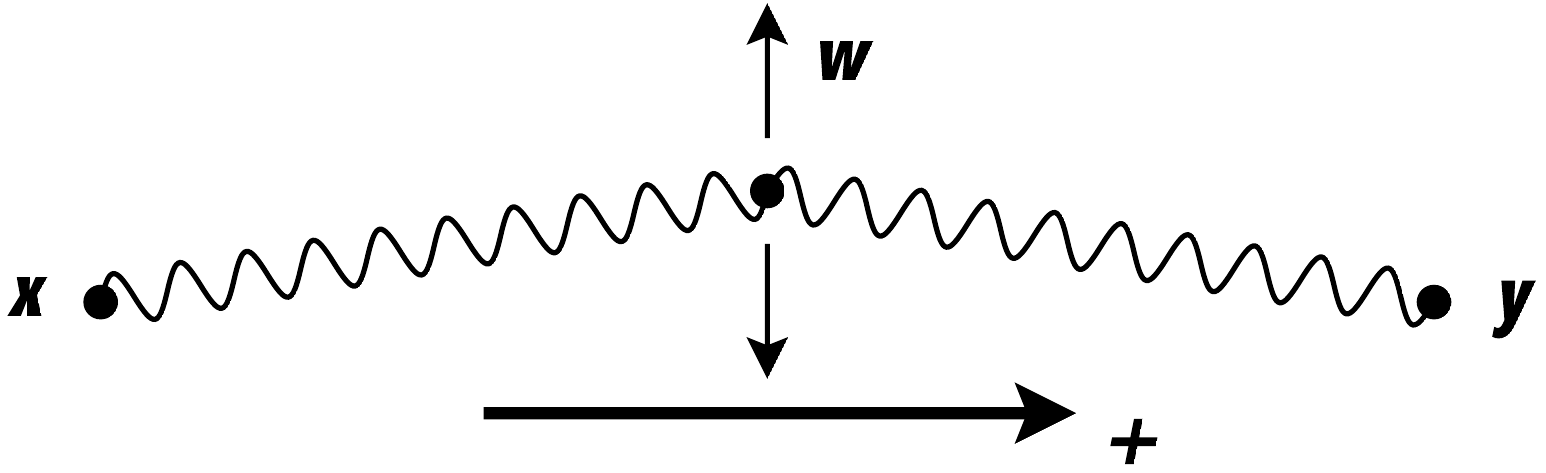}{6cm}}
\caption{One-vertex example. \label{fig:1vertex}}
\end{figure}

Once the theory is renormalized for ultraviolet divergences, Green function integrals in $G_N$ are singular only at pinches in the complex integrations over positions of vertices, $y_k$   between ``incoming" and ``outgoing" propagators (on the light cone) \cite{Date:1982un,Erdogan:2013bga}.
 An example is given by a single vertex between two lines,  as illustrated in Fig. \ref{fig:1vertex}.  The vertex at $w$ in the figure is connected to one or more additional lines in general.  Whether or not these additional lines are near the light cone, the two propagators shown in the figure provide the integral,
 \bea
&\ & \int d^D w\ \frac{1}{\left(-2(y-w)^+(y-w)^- + ({\bf y}_\perp-{\bf w}_\perp)^2+i\ep\right)^{1-\vep}}
\nn\\
&\ & \hspace{30mm} \times
 \frac{1}{\left(-2(w-x)^+(w-x)^- + ({\bf w}_\perp-{\bf x}_\perp)^2+i\ep\right)^{1-\vep}} \ , 
\label{eq:1-vertex-pinch}
\eea
which gives a pinch at $w^-,{\bf w}_\perp=0$ when $x$ and $y$ are lined up in the $+$ direction, with $y^+>w^+>x^+$.    This illustrates in coordinate space the momentum-space result that pinches correspond to physically-realizable scattering processes, in which on-shell particles follow classical paths between vertices.  
 
\begin{figure}[h]
\centerline{  \figscale{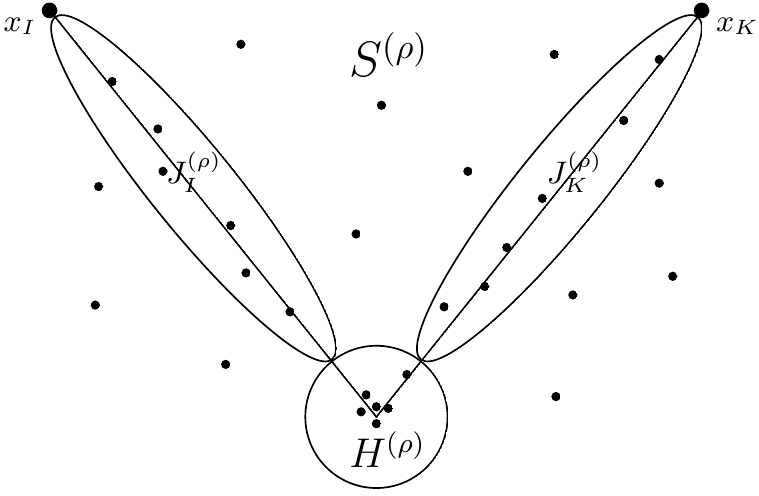}{6cm}}
\caption{Arrangements of vertices near a pinch surface with two jet subdiagrams, in the $\beta_I^\mu \propto x_I^\mu$ and $\beta_K^\mu \propto x_K^\mu$ directions. \label{fig:vertices-arranged}}
\end{figure}

\begin{figure}
\centerline{\figscale{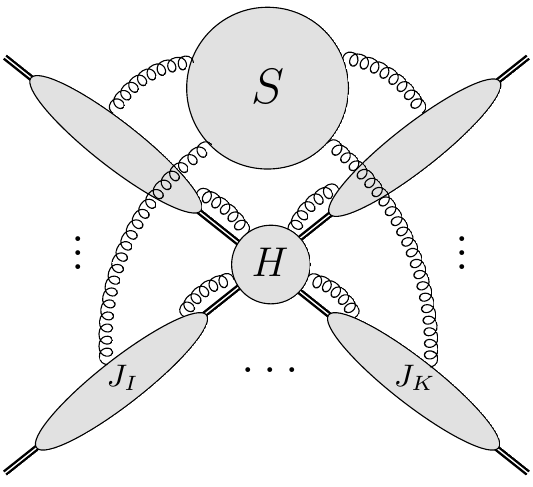}{4.5cm} }
\caption{General pinch surface  ($\rho$) in coordinate space.    \label{fig:gen-ps}}
\end{figure}

Applying this reasoning to higher orders, the general ``leading (-power) region" corresponds to the most general physical picture made up of on-shell massless lines, as illustrated by Figs.\ \ref{fig:vertices-arranged} and \ref{fig:gen-ps} \cite{Erdogan:2013bga,Erdogan:2014gha} for Eq.\ (\ref{eq:VEV}).   This is just the pattern found in momentum space for elastic scattering amplitudes \cite{Akhoury,Sen:1982bt}.   Figure \ref{fig:vertices-arranged} can be thought of as a two-dimensional projection of part of a physical scattering process.   Jets are in directions defined by velocity vectors, $\beta_I^\mu \propto x_I^\mu$, from the position of the hard scattering to external points $x^\mu_I$.  By translation invariance, the hard scattering may be fixed at the origin. Vertices cluster along the $\beta_I$, near the origin, or are at finite distances from these.  More specifically, Fig.\ \ref{fig:vertices-arranged} is a picture of ``where the vertices are",  when they are on or near a pinch surface, $\rho$: i) Vertices in $H^{(\rho)}$ are near the origin; ii) Vertices in $J_I^{(\rho)}$ are near rays $\beta_I^\mu$ for $x_I^2 \rightarrow 0$; iii) Vertices in $S^{(\rho)}$ are separated from the origin and the rays.  A global portrait of such a pinch surface is shown in Fig.\ \ref{fig:gen-ps} \cite{Akhoury,Sen:1982bt}.

We emphasize that a single picture like Fig.\ \ref{fig:vertices-arranged} or Fig.\ \ref{fig:gen-ps} represents the pinch surfaces of large numbers of diagrams, related by connecting the vertices of the picture in all possible ways.   The organization of pinch surfaces thus naturally incorporates much of the gauge symmetry of the theory, which is not realized on a diagram-by-diagram basis.   

\section{Approximations, subtractions and factorization}

 For each lightlike vector $\beta_I$, $\beta_I^2 \rightarrow 0$, we introduce an ``opposite-moving" lightlike vector, $\bbeta_I$, $\bar\beta_I^2=0, \ \beta_I\cdot\bar\beta_I=1$.    Then, for every diagram that corresponds to a leading region $\rho$, we introduce an approximation operator $t_\rho$ acting on each $n$th-order diagram $\gamma^{(n)}$ \cite{Erdogan:2014gha}, which isolates its divergent behavior in region $\rho$, indicated by subscript $\mathrm{div}[\rho]$,
\bea
t_\rho \gamma^{(n)}\ &=& \ \gamma^{(n)}\bigg |_{\mathrm{div}[\rho]}
\nonumber\\
&\ & \hspace{-20mm} \equiv\ 
\prod_I
\int d\tau^{(I)} \ S^{(\rho)}_{\{\mu_I\}}(\{\tau^{(I)}\})\ \beta_I^{\mu_I}\bar\beta_{I,\mu'_I}
\nn\\
&\ & \hspace{1mm} \times\ \int d\eta^{(I)}\;
\int d^{D-1} z^{(I)} \ J_I^{(\rho)\mu_I'\nu_I'}(z^{(I)},\eta^{(I)})\ \bar\beta_{I,\nu'_I}\beta_I^{\nu_I}\ \int d^{D-1} y^{(I)} \ H^{(\rho)}_{\{\nu_I\}}(y^{(I)}) \ . 
\label{eq:t-rho}
\eea
For gluons attaching the ``soft" function $S^{(\rho)}$ to jet $I$ in direction $\beta_I$, we keep only the $\bbeta_I$ polarization and the coordinates $\tau^{(I)}$ of vertices along the $\beta_I$ direction.
This approximation is closely related to one of the starting points for soft-collinear effective theories \cite{Bauer:2000yr}.   In its action on the jet $J_I$, Eq.\ (\ref{eq:t-rho}) is equivalent to the replacement \cite{Becher:2014oda}
\bea
A^\mu\ =\ A^\mu_c(x)\ +\ \bar n^\mu\, n\cdot A_s(\bar n \cdot x)\ ,
\eea
with $A_{c,s}$ the collinear and soft gluon fields, respectively, and $n^\mu$ ($\bar n^\mu$) playing the role of $\bar\beta^\mu_I$ ($\beta^\mu _I$).  

Constructed as in (\ref{eq:t-rho}), the operators $t_\rho$ organize all divergences as external points approach the light cone relative to the hard scattering:
\bea
 \left[ \gamma^{(n)}\ +\  \sum_{N}\  \prod_{\rho\in N} \big(-t_{\rho}\big)\, \gamma^{(n)}  \right]\, \bigg |_{{\rm div}[\rho]}
\  =  \ 0\, .
   \eea 
Note that each individual operator, $t_\rho$ acts to approximate the integrand by its leading behavior only near the singular surface $\rho$.   Following Collins and Soper \cite{Collins:1981uk} in axial gauge and Collins \cite{Collins:2011zzd}  in covariant gauge, the sum is over all possible sets $N$ of nested regions, which cancels overlapping divergences.  This procedure also formalizes a strategy of regions \cite{Beneke:1997zp}. 
Within each region $\rho$, only the term with approximation $t_\rho$ contributes, while all others cancel, but each approximation extends over all coordinate space.    

\begin{figure}
\centerline{\figscale{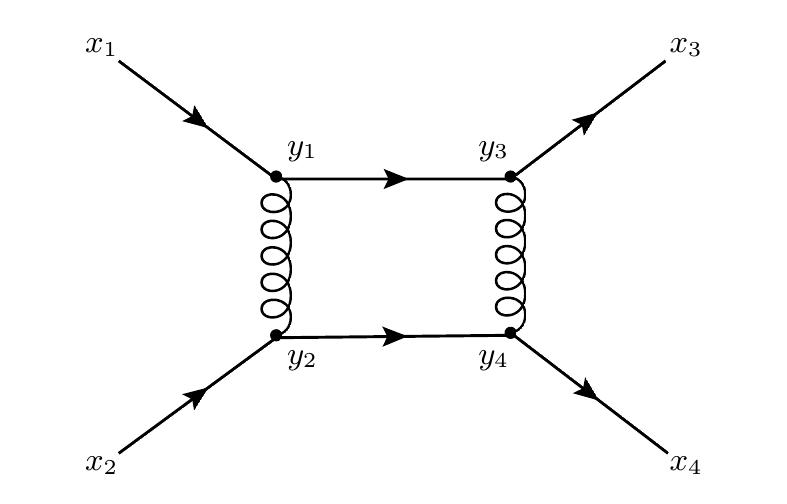}{6cm}}
\caption{Example with disjoint choices for the hard scattering.}
\label{fig:disjoint-hard}
\end{figure}

Already at one loop, nesting for fixed-angle scattering  becomes nontrivial, because in QCD hard scatterings can be disjoint, as illustrated by Fig.\ \ref{fig:disjoint-hard} for quark-quark scattering.  Either gluon in the figure can mediate the hard scattering, with the other gluon soft, collinear, or included in the hard scattering.  In any case, double counting can be avoided to all orders. 
The same arguments are applicable to momentum space, and the relation to the pattern observed in higher order exact calculations  is clear, where elaborate nested approximations are a familiar feature.      

   As $x_I^2\rightarrow 0$ relative to the hard scattering ($H$), jet vertices line up along intersecting light cones: $x_I^\mu \propto \beta^\mu_I$, $\beta_I^2=0$, and the subtraction formalism allows the derivation of a factorized amplitude in coordinate space.  
The result for vacuum expectation values, Eq.\ (\ref{eq:VEV}), with fixed-angle geometry is a factorized expression very much like that familiar in momentum space \cite{Erdogan:2013bga,Erdogan:2014gha},
\bea
G_N\left(\left\{x_I\right\}\right)\ &=&\  \prod_{I=1}^N\ \int d\eta_I\  j^{\rm part}_I(x_I,\eta_I\bar\beta_I)\  
S_{\rm ren}(\{\beta_I\cdot \beta_J\})\ \  H\left (\{\eta_I\bar\beta_I\} \right)\, ,
\label{eq:vev-fact}
\eea
with short-distance function $H$, and with ``jet" functions,
\bea
j_I^{\, \rm part}(x_I,\eta_I\bar\beta_I)\
=\ c_I(\beta_I,\bar\beta_I)
\bigg \langle 0\left| T\bigg(  \phi(x_I)\, \phi^\dagger(\eta_I\bar\beta_I) \Phi_{\bar\beta_I}^{[f_\phi]}{}^\dagger(\infty,\eta_I\bar\beta_I) \bigg)\right|0  \bigg \rangle\, ,
\label{eq:vev-j}
\eea
where $\Phi^{[f_\phi]}_{\bar\beta_I}$ is a Wilson line in the color representation $f_{\phi_I}$ of field $\phi$, and in the direction $\bar\beta_I$ opposite to the jet direction, while $c_I$ is a normalization factor \cite{Erdogan:2014gha}\@.   The soft function, $S_{\rm ren}$ is the vacuum expectation value of a product of Wilson lines in the lightlike $\beta_I$ directions. 

\section{Coordinate picture for cusps and polygons in QCD}

\begin{figure}
\centerline{\figscale{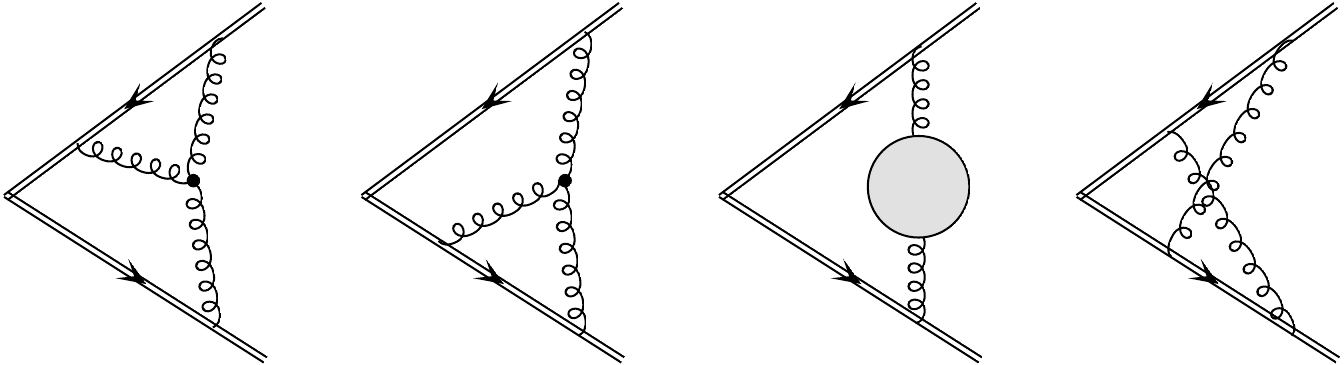}{10cm}}
\caption{Two-loop web diagrams for the cusp.  All have the same color factor, $C_AC_F$. }
 \label{fig:2-loop-webs}
 \end{figure}

In many hard-scattering amplitudes and cross sections, soft radiation is organized in terms of Wilson lines.    A coordinate treatment of Wilson lines makes possible a combinatoric proof of exponentiation in coordinate space \cite{Mitov:2010rp}.   The argument is more general, but here we'll consider just the cusp.   This will enable us to exhibit an interesting ``geometrical" interpretation directly in QCD, which for large numbers of colors, $N_c$, extends to arbitrary soft functions as in (\ref{eq:vev-fact}).   To be specific, the cusp, that is, the 2-line eikonal form factor joined at a singlet vertex, is the exponential of a sum of  ``webs"  \cite{Gatheral:1983cz},
\bea
A &=&  \exp\left[\sum_{i=1}^{\infty} w^{(i)} \right] \, .
\label{eq:A-exp-w}
\eea
The webs at $i$ loops, $w^{(i)}$, are two-eikonal irreducible diagrams, shown in Fig.\ \ref{fig:2-loop-webs} for two loops.  The contribution from each diagram in the figure is computed in terms of propagators in the usual way, in either momentum or coordinate space, but as webs in this case their color factors are all the same. For example, when the Wilson  lines are in the fundamental representation, the color factors are $C_F C_A$ only, with no $C_F^2$-terms.

Here's how it works.  Say we know the exponent $w^{(i)}$ to order $N$.  We then expand to $N+1$st order, as a sum of exponentiated web diagrams,\bea
A^{(N+1)} &=& \left(\, \exp\left[\sum_{i=1}^{N+1} w^{(i)} \right] \, \right)^{(N+1)}\, ,
\label{eq:A-N+1}
\eea
where the superscript instructs us to keep contributions only to the order indicated.
At the same time, $A^{(N+1)}$ is the sum of all $N+1$st-order diagrams,
\bea
A^{(N+1)} &=& \sum_{D^{(N+1)}}  D^{(N+1)} \, .
\label{eq:AasDsum}
\eea
Applied to Eq.\ (\ref{eq:A-exp-w}) for the exponentiated cusp, these two relations imply a formula for the highest order web that appears in the exponent,
\bea
w^{(N+1)} &=& 
\sum_{D^{(N+1)}}  D^{(N+1)} - \left[\, \sum_{m=2}^{N+1} \frac{1}{m!}
\sum_{i_m=1}^N \dots \sum_{i_1=1}^N\,
w^{(i_m)}\, w^{(i_{m-1})} \dots w^{(i_1)} \right]^{(N+1)}\, .
\label{eq:w-N+1}
\eea	
The simplest  (and most general) constructions of webs are in coordinate space, and start with this relation \cite{Mitov:2010rp}. 

An important property of webs that can be proved in this way is that they lack soft or collinear subdivergences \cite{Erdogan:2014gha,Gatheral:1983cz,Erdogan:2011yc}.   The webs  of a given order automatically combine to cancel all divergences from singular regions where a proper subdiagram of a web is collinear to either one of the Wilson lines.  Divergences arise only when every line in each web diagram is either hard, soft, or collinear to one Wilson line or the other.  In this way,  a web acts like a single gluon.

In coordinate space with an infrared cutoff $L$, we can then write the cusp exponent (the sum of webs) as
\bea
E(L,\vep)
= 
\int_0^L \frac{d\lambda}{\lambda}\, 
 \frac{d\sigma}{\sigma}\, 
w \left(\as\left(1/\lambda\sigma\right)\right) \ ,
\eea
in terms of a function of the running coupling only, with
$\sigma$ and $\lambda$ the maximal distances from the origin at which the web attaches along the eikonal lines.   The invariant size of the web fixes the running coupling, and the function $w$ is given by
\bea
w = -\ \frac{1}{2} \Gamma_{\rm cusp}(\as) + {\cal O}(\vep)\, ,
\eea
where $\Gamma_{\rm cusp}=\left(\frac{\alpha_s}{\pi}\right) C_a\left[1+\left(\frac{\alpha_s}{\pi}\right)\left( \left(\frac{67}{36}-\frac{\pi^2}{12}\right)C_A\,-\,\frac{5}{18}n_fT_f\right)+\dots\right]$ is the familiar cusp anomalous dimension.
In QCD the coupling runs as the integral passes over the $x_0,x_3$ plane at $x_\perp=0$, and if we identify the distance in an additional dimension with the argument of the running coupling, these distances define a minimal surface in  a 5-dimensional space,  just as  in the strong-coupling limit for conformal theories \cite{Alday:2007hr}, but here for any coupling. 

\begin{figure}
\centerline{\figscale{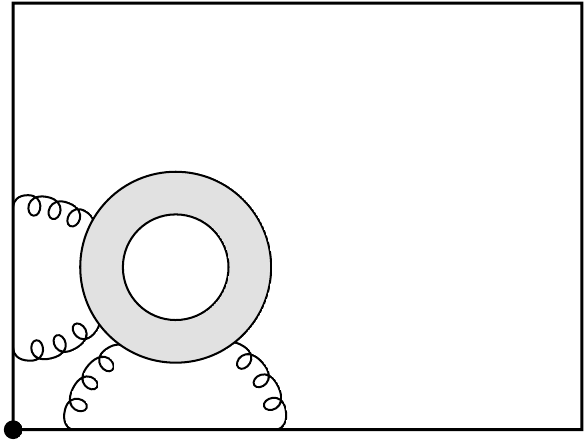}{3cm}}
\label{fig:web2}
\caption{A web at the corner of a polygon. \label{fig:cusp-poly}}
\end{figure}

There is another interesting correspondence in polygonal Wilson lines \cite{Korchemskaya:1992je,Drummond:2007aua}, now neglecting the running of the coupling.   Webs appear at the corners of a polygon, as in Fig.\ \ref{fig:cusp-poly}, and when defined in a gauge-invariant fashion give the leading singularities. 
Neglecting the running of $\as$, we find another analogy to minimal surfaces in five dimensions \cite{Erdogan:2011yc},
\bea
\sum_{a=1}^4 W_a(\beta_a,\beta'_a)
&=&  \int_{-1}^1 dy_1\, \int_{-1}^1 dy_2\,
\frac{4w}
{(1-y_1^2)(1-y_2^2)}\, ,
\eea
with $W_a$ the web at corner $a$\@.  
Again, the result is as found in Ref.\ \cite{Alday:2007hr}\@.

\section{A coordinate picture for cross sections } 

Many of the coordinate-space considerations of Refs.\ \cite{Erdogan:2013bga} and \cite{Erdogan:2014gha} can be extended to cross sections. Here, we will ask only the most basic questions, how infrared singularities occur in coordinate integrals and how they cancel in fully inclusive cross sections.    Schematically, the contribution of a specific final state to a fully inclusive cross section can be represented as 
\bea 
\prod_j \int \frac{d^Dk_j}{(2\pi)^{D}}\  (2\pi)\, \delta_+\left (k_j^2\right )\ A^*(\{k_j\})\, (2\pi)^D\,\delta^D\left (q-\sum_{k\in \{k_j\}} k \right)\  \ A(\{k_j\})\, , \label{eq:cross-sec}
\eea
with $A(\{k_j\})$ the amplitude for the production of a final state with particles with momenta $\{k_j\}$.    For this argument, we suppress dependence on the initial state.  

To express this cross section as a coordinate integral, we need in addition to Eq.\ (\ref{eq:scalarprop}) the Fourier transform of the ``cut propagator" with momentum flowing out of a vertex at point $w$ in the amplitude and into point $y$ in the complex conjugate, 
\bea
\int \frac{d^D k}{(2\pi)^D}\ \e^{-ik\cdot (y-w) }\, (2\pi)\, \delta_+\left( k^2\right) \ =\ \frac{\Gamma(1-\vep)}{4\pi^{2-\vep}}\ \frac{1}{(-(y-w)^2+ i\ep (y^0-w^0))^{1-\vep}}\, .
\label{eq:cut-prop}
\eea
For $y^0>w^0$, this is equal to a propagator in the amplitude, and for $w^0>y^0$, it equals a propagator in the complex conjugate amplitude.   This feature leads to pinch surfaces for cut diagrams, and hence in cross sections.   
\begin{figure}
\centerline{ \figscale{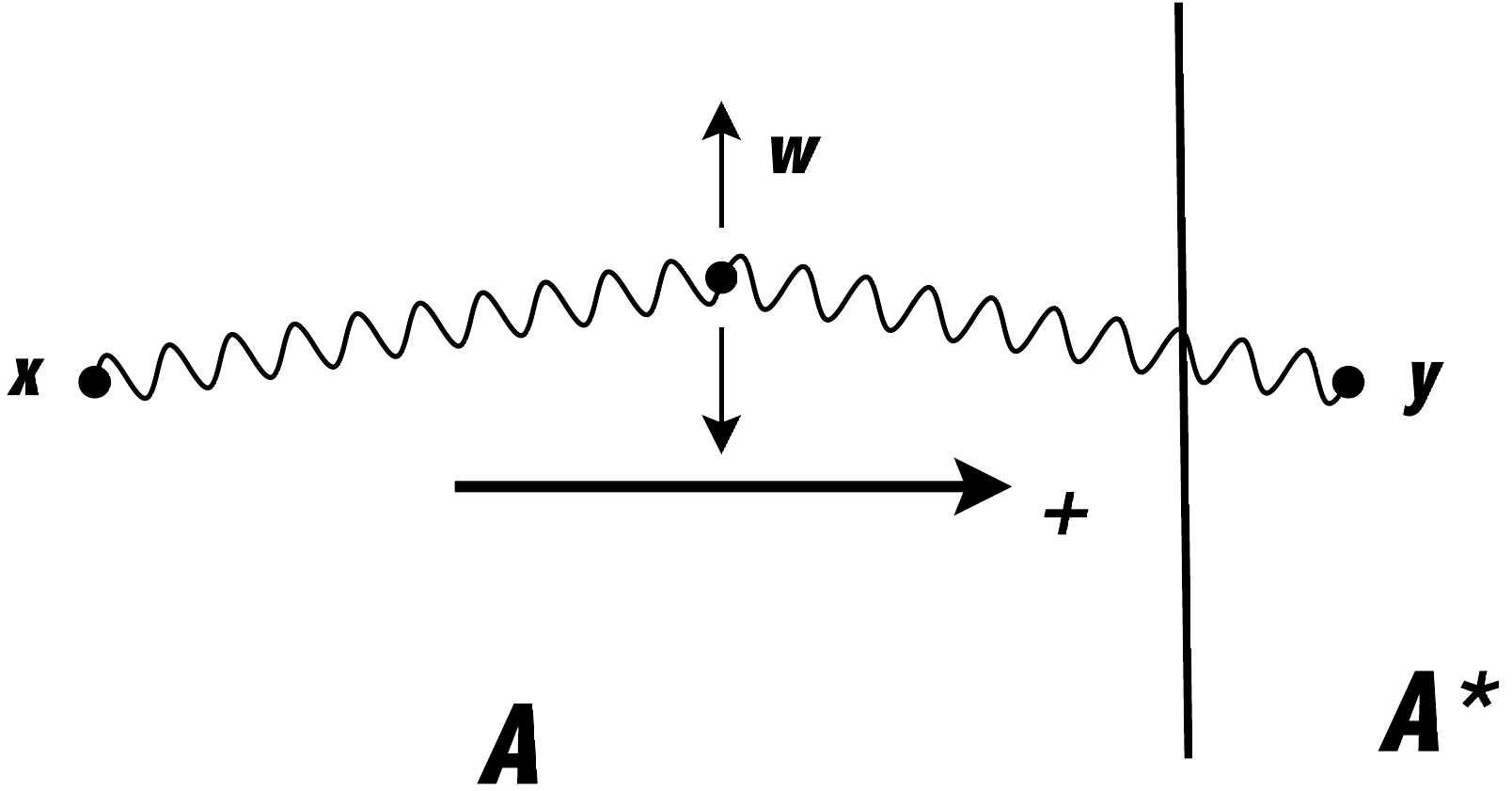}{5cm}}
\caption{Illustration of the role of the cut propagator  in cross sections. \label{fig:x-w-y-xsec}}
\end{figure}
In Fig.\ \ref{fig:x-w-y-xsec}, consider a collinear configuration of vertices $x$, $w$, and $y$ near the $x^-=0$ light cone.  Then, assuming $x$ and $y$ are pinched on the light cone in the $+$ direction ($x^-=x_\perp=y^-=y_\perp =0$), $w$ is also pinched in the plus direction, as follows.  To have a pinch at vertex $x$, we know from the analysis above that $w^0>x^0$.  Now if $y^0>w^0$, the ``cut" propagator (\ref{eq:cut-prop}) has  imaginary part $\vep(y^0-w^0)>0$ in the denominator.   When the points line up on the light cone the cut denominator has a pole at $w^-=-i\ep(y^0-w^0)/2(y^+-w^+)$, which is in the lower half-plane because on the light cone plus components are ordered in the same way as time components.  This is the same arrangement as if vertex $y$ were in the amplitude, pinching with the  $w^-$ pole from the $w-x$ propagator, which is in the upper half-plane.  If $w^0>y^0$, the cut propagator has imaginary part $\vep(y^0-w^0)<0$, but because now $y^+-w^+<0$, the pole remains in the lower half plane and continues to give a pinch in the $w^-$ integral on the light cone.

All this means that cross sections have the same pinch surfaces as amplitudes,  characterized by the same types of physical processes, but with energy flow reversed in the complex conjugate.  

Because vertices of both $A$ and $A^*$ are ordered at any pinch surface, the vertex with the largest time is always adjacent to the final state, and connected across the cut.  This leads to a picture of how infrared divergences occur for individual final states through integrals over the positions of vertices, and also how cancellation can take place in the sum over final states.   

\begin{figure}
\centerline{ \figscale{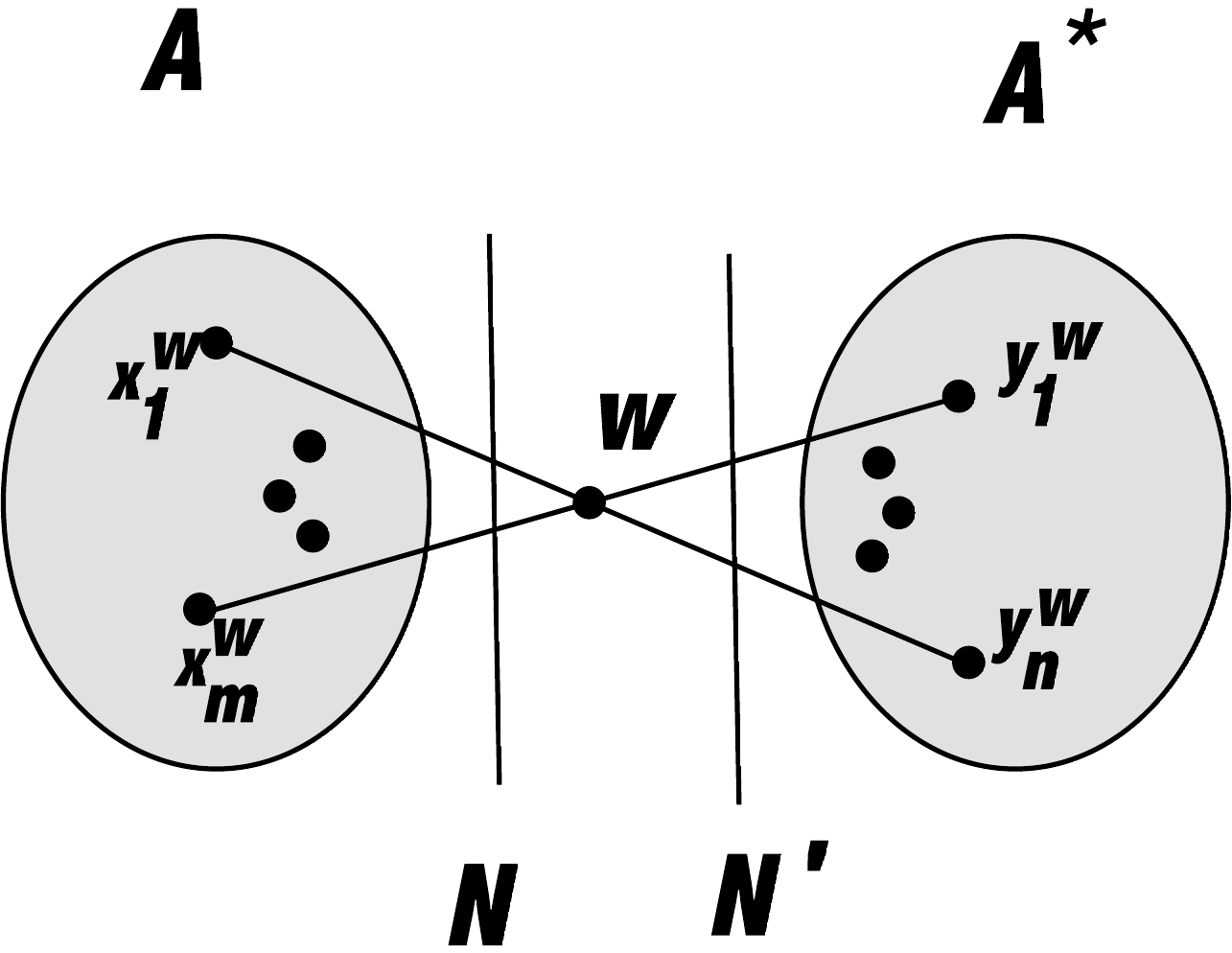}{4cm}}
\caption{Schematic arrangement for the vertex of ``largest time".\label{fig:states}}
\end{figure}

The cancellation of IR divergences in an inclusive cross section depends directly on  the hermiticity of the interaction.  In Fig.\ \ref{fig:states}, let $w^0$ be the ``largest time" \cite{Veltman:1963th} of any interaction vertex in either the amplitude or its complex conjugate; then certainly
\bea
w^0\ >\ x_i^0 \quad {\rm and} \quad w^0 \ >\ y_j^0
\eea
for all $i$ and $j$.
Vertex $w$ may be in $A$ or in $A^*$, and we must sum over both cuts, $N$ and $N'$, for which vertex $w$ is in $A^*$ and $A$, respectively.
Schematically, we can represent this combination, keeping only the relevant denominators, as the sum of two terms
\bea
&\ &  \int d^Dw\ \Bigg \{\ \ \prod_j \frac{1}{(-(y^w_j -w)^2 - i\ep)^{1-\vep}}\ \left [ -iV(\partial_w)\right ]\ \prod_i \frac{1}{(-(w-x^w_i)^2 + i\ep(w-x^w_i)^0 )^{1-\vep}}
 \nn
\\
&\ & \hspace{5mm} +\ \prod_j \frac{1}{( -(y^w_j -w)^2 + i\ep(y^w_j-w)^0)^{1-\vep}}\ \left[ iV(\partial_w)\right]\ \prod_i \frac{1}{(-(w-x^w_i)^2 + i\ep)^{1-\vep}} \Bigg \}\, ,
\eea
for $w$ in $A^*$ and $A$, respectively.
Typically, of course, the number of lines attached to vertex $w$ on either side of the cut is one or two.   In any case, infrared divergences may develop when any pair from $x^w_i$ and $y^w_j$ are collinear to $w$  on the light cone, which produces a pinch, and/or when $w$ becomes large.  In either of these cases, the sum vanishes because $w^0$ is the largest time.   This is the way in which infrared divergences cancel in coordinate integrals for the inclusive cross section.   
\section*{Some final thoughts}

The coordinate picture we have just described offers an additional viewpoint into the dynamics of hard scattering.  Considerably more work will be required to come to a full understanding of the cancellation of infrared sensitivity in semi-inclusive cross sections, and hence to factorization of physical observables from a coordinate viewpoint.  It is possible that this approach will lead to further insights into the variety of factorizations and the limitations of each.   It is even possible that a coordinate point of view may make it possible to combine a weak coupling, perturbative viewpoint with nonperturbative, emergent degrees of freedom.  The results sketched above also offer progress in systematizing all-orders factorization for QCD hard scattering, both in coordinate and momentum space.

\acknowledgments

We thank the organizers of Radcor 2015 and Loopfest XIV for the invitation to present our work.  This work was supported in part by the National Science Foundation,  grants No.\ PHY-0969739 and No.\ PHY-1316617. The work of O.E. was also supported by the Isaac Newton Trust and DOE contracts DOE-ER-40682-143 and DEACO2-C6H03000.

\end{document}